\documentclass[12pt]{iopart}
\usepackage{epsfig,amssymb,psfig}

\def\spose#1{\hbox to 0pt{#1\hss}}
\def\lesssim{\mathrel{\spose{\lower 3pt\hbox{$\mathchar"218$}}
 \raise 2.0pt\hbox{$\mathchar"13C$}}}
\def\gtrsim{\mathrel{\spose{\lower 3pt\hbox{$\mathchar"218$}}
 \raise 2.0pt\hbox{$\mathchar"13E$}}}

\def\<{\langle}
\def\>{\rangle}
\newcommand\onlinecite{\cite}
\newcommand{\mvec}[1]{\ensuremath{{{#1}}}}

\begin{document}

\title{ 
The uniformly frustrated two-dimensional $XY$ model in the limit of weak frustration
}

\author{Vincenzo Alba$^1$, Andrea Pelissetto$^2$ and Ettore Vicari$^3$}
\address{$^1$ Scuola Normale Superiore and INFN, I-56126 Pisa, Italy}
\address{$^2$ Dipartimento di Fisica dell'Universit\`a di Roma ``La Sapienza"
        and INFN, Sezione di Roma I, I-00185 Roma, Italy}
\address{$^3$ Dipartimento di Fisica dell'Universit\`a di Pisa
        and INFN, Sezione di Pisa, I-56127 Pisa, Italy}

\ead{
Andrea.Pelissetto@roma1.infn.it,
Ettore.Vicari@df.unipi.it}

\begin{abstract}
We consider the two-dimensional uniformly frustrated $XY$ model in the 
limit of small frustration, which is equivalent to an $XY$ system, 
for instance a Josephson junction array, in a weak uniform magnetic field 
applied along a direction orthogonal to the lattice. We show that the 
uniform frustration (equivalently, the magnetic field)
destabilizes the line of fixed points which characterize
the critical behaviour of the $XY$ model for $T \le T_{KT}$, where $T_{KT}$
is the Kosterlitz-Thouless transition temperature: the system is 
paramagnetic at any temperature for sufficiently small frustration.
We predict the critical behaviour of the correlation length and of 
gauge-invariant magnetic susceptibilities as the frustration goes to zero.
These predictions are fully confirmed by the numerical simulations.
\end{abstract}


\maketitle


\section{Introduction}

The uniformly frustrated two-dimensional (2D) $XY$ model is
defined by the lattice Hamiltonian
\begin{equation}
{\cal H} = - \sum_{\langle xy \rangle } {\rm Re} \,\psi_x U_{xy} \psi_y^*
= -  \sum_{\langle xy \rangle} {\rm cos}(\theta_x - \theta_y+A_{xy}),
\label{xymodB}
\end{equation}
where $\psi_x\equiv e^{i\theta_x}$ and $U_{xy}\equiv e^{i A_{xy}}$. 2D
arrays of coupled Josephson junctions in a magnetic field are interesting
physical realizations of this model \cite{FZ-01}. 
In this case, the sum $C(P_{nm})$ 
of the variables $A_{xy}$ along the links of an elementary plaquette $P_{nm}$,
\begin{eqnarray}
C(P_{nm}) & \equiv  &
A_{(n,m),(n+1,m)} + A_{(n+1,m),(n+1,m+1)} \nonumber \\
&& - 
 A_{(n,m+1),(n+1,m+1)} - A_{(n,m),(n,m+1)},
\end{eqnarray}
is related to the flux of an external magnetic field
applied along an orthogonal direction: $C(P_{nm}) = {a^2 B/\Phi_0}$,
where $a$ is the lattice spacing, $B$ is the magnetic field and
$2\Phi_0=hc/e$. Hamiltonian (\ref{xymodB}) depends on $A_{xy}$
through the phases $U_{xy}$ and thus the relevant physical quantity 
is the product of the phases around a plaquette, i.e.,
$U(P) \equiv \exp[i C(P)]$. If $U(P)$ is not 1, ${\cal H}$ is frustrated. 
In this paper we assume $U(P)$ to be 
independent of the chosen plaquette, i.e., that
\begin{equation}
   U(P) = e^{2\pi i f},
\label{deff}
\end{equation}
with $0\le f \le 1$, independent of $P$. Using the 
invariance of the Hamiltonian under the transformation $\psi_x\to\psi^*_x$,
it is not restrictive to take $f$ in the interval $0\le f \le 1/2$.
We will work in a finite lattice of size $L^2$ 
with periodic boundary conditions. Therefore, we have
\begin{equation}
  \prod_P U(P) = 1,
\end{equation}
where the product is extended over all lattice plaquettes. This implies that 
$f L^2$ must be an integer.

Hamiltonian (\ref{xymodB}) is invariant under the local gauge transformations
\begin{equation}
\psi_x \rightarrow V_x \psi_x, \qquad  U_{xy} \rightarrow V_x^* U_{xy} V_y ,
\label{ginv}
\end{equation}
where $V_x$ is a phase, $|V_x| = 1$.  Physical observables
must be gauge invariant. For such observables, the choice of the fields
$A_{xy}$ is irrelevant: only the value of $f$ is relevant.
In a finite volume, 
this statement is strictly true only if free boundary conditions are 
taken. If one considers periodic boundary conditions, one must also
specify the value of $\exp (i \sum A_{xy})$ along two non-trivial lattice paths
that wind around the lattice (they are sometimes called Polyakov loops). 
For instance, one must also fix
$P_1(m)  = \exp(i\sum_n A_{(n,m),(n+1,m)})$ and 
$P_2(m)  = \exp(i\sum_n A_{(m,n),(m,n+1)})$ for some fixed value of $m$. 
If we require the absence of magnetic circulation along these 
non-trivial paths, we must have $P_1(m) = P_2(m) = 1$ for any $m$. On a finite 
lattice of size $L^2$, this condition can be satisfied only if $f L$ is 
an integer, a condition that will be always satisfied in the numerical 
simulations that we shall present.

The critical behaviour of uniformly frustrated $XY$ models changes
dramatically with $f$.  For $f=0$ the model corresponds to 
the standard $XY$ model, which is not frustrated. It shows
a Kosterlitz-Thouless transition at $T_{KT}$ [on a square
lattice~\cite{Has-05} $T_{KT}=0.89294(8)$], where the correlation length
$\xi$ diverges as ${\rm ln}
\xi\sim (T-T_{KT})^{-1/2}$ for $T\gtrsim T_{KT}$; the low-temperature phase,
$T<T_{KT}$, is characterized by quasi long-range order---correlation functions 
decay algebraically---associated with a line of fixed points. In the 
case of maximal frustration, i.e. for $f=1/2$, the system undergoes two very
close continuous transitions (their critical temperature is
$T\approx 0.45$ on the square
lattice), respectively in the Ising and Kosterlitz-Thouless
universality classes, see, e.g., \onlinecite{HPV-05,Korshunov-06} 
and references therein. The critical behaviour for other values of $f$ is 
even more complex, see, e.g.,
\onlinecite{CD-85,KVB-95,FT-95,HW-95,LL-95,SMK-97,CS-85,PCKJC-00},
and \onlinecite{Ling-etal-96} for experiments.  There may be 
several transitions, whose nature
is not clear in most of the cases. Even the structure of the ground state is 
only partially understood \cite{TJ-83,SB-93,LLK-02}. For 
$f=1/n$, where $n$ is an integer number, if $T_c$ is the critical 
temperature where the paramagnetic phase ends, $T_c$ 
decreases with increasing $n$; for
example, \cite{LL-95} $T_c\lesssim 0.22$ for $f=1/3$ and~\cite{HW-95}
$T_c\lesssim 0.05,\,0.03$ for $n=30$ and 56, respectively.  These studies
suggest that $T_c$ vanishes~\cite{HW-95,FT-95} as $T_c\sim 1/n$ when
$n\to \infty$.  The critical behaviour for irrational values of $f$ 
is even less clear, see, e.g., 
\onlinecite{CS-85,PCKJC-00}. In this case, there are some indications
that the system is paramagnetic for any $T$ and that a glassy
transition occurs at zero temperature \cite{PCKJC-00}.

The above-mentioned works studied the critical behaviour
as a function of the temperature $T$, while
keeping the uniform frustration $f$ fixed. In this paper we investigate a
different critical limit, i.e. we consider the limit 
$f\to 0$ at fixed $T$ in the region $T\le T_{KT}$. 
In other words, we investigate 
the effect of a small uniform frustration 
on the low-temperature $XY$ critical
behaviour.  We show that a uniform frustration is a
relevant perturbation at the fixed points that occur in the 
$XY$ model for $T \le
T_{KT}$. As soon as $f$ is non-vanishing, the correlation length becomes finite
and the system is paramagnetic.

The critical behaviour for small values of $f$ can be understood within the 
Coulomb-gas picture \cite{FHS-78}. If one considers the Villain Hamiltonian
corresponding to (\ref{xymodB}), one can write the partition function
as 
\begin{equation}
Z_{\rm Villain} = \int \prod_{x} d\theta_x\, e^{-\beta {\cal H}} = Z_{SW} 
        \sum_{\{n_x\}} \exp\left(2 \pi \beta {\cal H}_{\rm CG}\right),
\label{Villain}
\end{equation}
where \cite{FHS-78} $Z_{SW}$ is the spin-wave contribution 
and ${\cal H}_{\rm CG}$ is the Coulomb-gas Hamiltonian:
\begin{equation}
{\cal H}_{\rm CG} = {1\over 2} 
\sum_{ij} (n_i-f) V(\mvec{r}_i-\mvec{r}_j) (n_j-f),
\label{cgas}
\end{equation}
where $n_i$ is an integer (vorticity) defined at the site $i$ of the dual
lattice and $V(\mvec{r})$ is the lattice Coulomb potential.
In (\ref{Villain}) the sum
over $n_x$ is restricted to configurations satisfying the
neutrality condition~\cite{FHS-78} $\sum_i (n_i-f)=0$.
For $f = 0$ and $T< T_{KT}$ this representation allows one to 
show that correlations functions decay algebraically. The two-point correlation function is the product of a spin-wave contribution, which decays algebraically,
and of a vortex contribution. For $T< T_{KT}$ charged vortices 
are strictly bound to form
dipoles and the corresponding correlation function also decays 
algebraically \cite{JKKN-77}. For $f>0$ the picture changes. 
For small $f$, in the temperature interval $f T_{KT} < T < T_{KT}$, there are 
unbounded particles with $n = 0$ and charge $-f$, which screen the 
Coulomb interaction among the vortices of charge $n - f\approx n$, $n\not=0$.
The Debye screening length can be easily computed. Consider a 
vortex of charge 1, surrounded by particles of charge $-f$. Since there 
is one charge $-f$ for each lattice site, complete screening is achieved 
when these charges occupy a circle of area $A$, such that $Af =  1$. 
Thus, the screening length $\xi$ should be proportional to $f^{-1/2}$. 
In this picture, for $f\to 0$, the system is equivalent to a dilute gas 
(the density is proportional to $f^{1/2}$) of neutral particles interacting
by means of a screened Coulomb potential $V_{sc}(r)$. We can thus 
perform a standard virial expansion to predict that the vortex-vortex 
correlation function is proportional to $V_{sc}(r)$, hence decays 
exponentially with a rate controlled by the Debye screening length.
This argument indicates that, for sufficiently small $f$ and any $T<T_{KT}$, 
the system is paramagnetic with a correlation length that scales as
\begin{equation}
\xi \sim f^{-1/2}, 
\label{xif}
\end{equation}
for $f\to 0$. 

Equation (\ref{xif}) can also be predicted by simple dimensional 
arguments. For a given value of $f$ and $T$, consider a 
real-space renormalisation-group (RG) transformation. Eliminate lattice sites
obtaining a lattice with a link length that is twice that of the original 
lattice. In lattice units we have
$\xi' = \xi/2$, where we use a prime for quantities that refer to the 
decimated lattice. Analogously, we obtain $f' = 4 f$ for the frustration 
parameter. 
It follows $\xi' {f'}^{1/2} = \xi {f}^{1/2}$. This quantity is therefore
constant under RG transformations, i.e. $\xi {f}^{1/2} = c$.
Under the RG transformation, the Hamiltonian parameters also change.
In particular, the transformation induces a temperature change
$T\to T'$. However, for small $f$, one is close to the $XY$ line 
of fixed points and thus we expect $T'\approx T$. Thus, the 
condition $\xi f^{1/2} = c$ holds at (approximately) fixed temperature
and $f\to 0$.
Therefore, it implies (\ref{xif}).

In this paper we wish to verify numerically (\ref{xif}) and 
study the critical behaviour of gauge-invariant susceptibilities (they will be
defined in the next section). Note that,
in a sense, at fixed $T\le T_{KT}$, the magnetic flux $f$
plays the role of the reduced temperature, with an associated
correlation-length exponent $\nu=1/2$.

The paper is organised as follows. In Sec.~\ref{sec2} we define 
gauge-invariant
correlation functions, the associated susceptibilities and correlation 
lengths, and discuss the expected critical behaviour. In Sec.~\ref{sec3}
we present some Monte Carlo (MC) results that fully confirm the 
theoretical predictions.

\section{Definitions and general scaling properties} \label{sec2}

In order to check prediction (\ref{xif}), we consider two different
gauge-invariant correlation functions:
\begin{eqnarray}
G_{sq}(\mvec{x};\mvec{y}) &\equiv &
   |\langle \psi_{\mvec{x}} \psi_{\mvec{y}}^* \rangle|^2,
\nonumber \\
G_\Gamma(\mvec{x};\mvec{y}) &\equiv& \langle {\rm Re} \, \psi_{\mvec{x}}
U[\Gamma_{\mvec{x};\mvec{y}}] \psi_{\mvec{y}}^* \rangle.
\label{gxdef}
\end{eqnarray}
Here $\Gamma_{\mvec{x};\mvec{y}}$ is a path that connects 
sites $\mvec{x}$ and $\mvec{y}$ and $U[\Gamma_{\mvec{x};\mvec{y}}]$ is
a product of phases associated with the links that 
belong to $\Gamma_{\mvec{x};\mvec{y}}$.
More precisely, 
if a link $\langle wz\rangle$ belongs to the 
path, $w$ and $z$ have coordinates $w = (w_1,w_2)$ and $z = (z_1,z_2)$,
such that $z_1 - w_1 \ge 0$ and $z_2 - w_2 \ge 0$, we define
$R_{wz} = U_{wz}$ if point $w$ occurs before point $z$ while 
moving along the path; otherwise, we set $R_{wz} = U_{wz}^*$.
The phase $U[\Gamma_{\mvec{x};\mvec{y}}]$ is the product of all the 
phases $R_{wz}$ associated with the links belonging to the path.

The definition (\ref{gxdef}) of $G_\Gamma(\mvec{x};\mvec{y})$ 
depends on a family of paths $\Gamma = \{\Gamma_{x;y}\}$. We assume this 
family to be translationally invariant: the path $\Gamma_{x;y}$ is obtained 
by rigidly translating the path $\Gamma_{0;y-x}$ that connects 
the origin to $y-x$. In this case, the correlation function 
$G_\Gamma(\mvec{x};\mvec{y})$ is uniquely
defined by specifying the paths from the origin to any point $x$.

Because of the presence of the gauge field, the Hamiltonian is 
not translationally invariant, nor is it symmetric under 
the symmetry transformations of the lattice. Nonetheless, there 
are generalized symmetries of the Hamiltonian 
that also involve gauge transformations. 
For instance, if $Lf$ is an integer, the Hamiltonian is invariant 
under the generalized translations
\begin{eqnarray}
\psi'_{(n,m)} = \psi_{(n+1,m)} U_{(n,m),(n+1,m)}^* e^{-2 \pi i m f}, 
\nonumber \\
\psi'_{(n,m)} = \psi_{(n,m+1)} U_{(n,m),(n,m+1)}^* e^{2 \pi i n f}.
\label{transl}
\end{eqnarray}
Gauge-invariant correlation functions are invariant under these 
transformations. This implies that they do not depend on $x$ and $y$
separately, but only on the difference $y - x$.
This invariance can be understood intuitively if one notes that 
gauge-invariant quantities
should only depend on the value of the flux through a plaquette, i.e.,
$U(P)$, and of the Polyakov correlations $P_1(m)$ and $P_2(m)$. 
In our model $U(P)$ is independent of $P$ and, if $Lf$ is an
integer, $P_1(m)$ and $P_2(m)$ do not depend on $m$: hence,
translation invariance holds. 

Analogously, the Hamiltonian is invariant under generalized transformations 
that involve lattice symmetries and gauge transformations.
For instance, in infinite volume the Hamiltonian is invariant under the 
generalized reflection transformations 
\begin{eqnarray}
\psi'_{(n,m)} &=& \psi_{(-n,m)}^* K_m^* \prod_{k=0}^{|n|-1} [U_{(k,m),(k+1,m)}
        U^*_{(-k-1,m),(-k,m)}], 
\end{eqnarray}
where
\begin{equation}
K_m = \cases{\displaystyle{
     \prod_{k=0}^{m-1} U^2_{(0,k),(0,k+1)}} & for $m \ge 1$, \cr
     1                                      & for $m=0$, \cr
     \displaystyle{
     \prod_{k=0}^{-m-1} U^{*2}_{(0,k+m),(0,k+m+1)}} &
                                       for $m \le -1$.
     }
\end{equation}
Under these symmetries $G_{sq}(x;y)$ transforms covariantly. If $T$ is 
a lattice symmetry, $G_{sq}(x;y) = G_{sq}(Tx;Ty)$. These relations do not hold
in general for $G_\Gamma(x;y)$ since a lattice symmetry also changes the 
path family.

Given $G_\Gamma(x;y)$ and $G_{sq}(x;y)$, 
we define the corresponding susceptibilities
\begin{eqnarray}
\chi_\Gamma \equiv  \sum_y G_\Gamma(x;y), \qquad\qquad 
\chi_{sq}  \equiv \sum_y G_{sq}(x;y),
\label{chi-def}
\end{eqnarray}
where the sums are extended over all lattice points $y$. Because of 
translational invariance, $\chi_{sq}$ and $\chi_\Gamma$ do not depend
on the point $x$. Of course, $\chi_\Gamma$ depends
on the family of paths $\Gamma = \{\Gamma_{x;y}\}$. 
Then, for any gauge-invariant correlation function $G(x;y)$ 
we define on a finite lattice of size $L^2$ 
\begin{eqnarray}
F \equiv  \sum_{y\equiv (y_1,y_2)} \cos [q_{\rm min} (y_1-x_1)] G(x;y)
\label{def-Fx}
\end{eqnarray}
where $x \equiv  (x_1,x_2) $ and $q_{\rm min} \equiv  2\pi/L$. 
The correlation length is defined by 
\begin{equation}
\xi^2 \equiv  {1\over 4 \sin^2(q_{\rm min}/2)} {\chi - F\over F}.
\qquad\qquad
\end{equation}
Note that an equally good definition of $F$ is
\begin{eqnarray}
F \equiv  \sum_{y\equiv (y_1,y_2)} \cos [q_{\rm min} (y_2-x_2)] G(x;y).
\end{eqnarray}
For the correlation function $G_{sq}(x;y)$, one can show that these two 
definitions of $F$ are equivalent, but this is not generically 
the case of $G_\Gamma(x;y)$,
since this quantity is not symmetric under lattice transformations.
In the following we use definition (\ref{def-Fx}) for $F$.

In the introduction we derived a prediction for the correlation length,
$\xi \sim f^{-1/2}$. 
We wish now to obtain a similar result for the susceptibilities.
In order to predict their scaling behaviour,
let us note that, for $f = 0$ and $T\le T_{KT}$, 
$\langle \psi_0\psi_x^*\rangle$ decays algebraically, 
i.e., $\langle \psi_0\psi_x^*\rangle \sim x^{-\eta(T)}$. The critical 
exponent $\eta(T)$ depends on $T$ and varies between $\eta(0) = 0$ and 
$\eta(T_{KT})=1/4$. For $f\not = 0$, it is natural to assume that 
\begin{eqnarray}
\chi_\Gamma \sim \int_{x < \xi} d^2x\, x^{-\eta(T)} \sim
\xi^{2-\eta(T)} \sim f^{-1+\eta(T)/2}, \nonumber \\
\chi_{sq} \sim \int_{x < \xi} d^2x\, x^{-2\eta(T)} \sim
\xi^{2-2\eta(T)} \sim f^{-1+\eta(T)}. 
\label{chibeh}
\end{eqnarray}
In particular, these equations predict $\chi_\Gamma \sim f^{-7/8}$ 
and $\chi_{sq} \sim f^{-3/4}$ at $T = T_{KT}$.

The check of the previous prediction for $\chi_{sq}$ does not present 
conceptual difficulties. Instead, when considering $\chi_\Gamma$, one shoud
keep in mind that this quantity depends on a path family. 
Thus, there is a natural question that should be considered first.
Given a path family $\Gamma^{(f_1)}$ for a given value $f=f_1$ of the 
frustration parameter, we must specify which path family
$\Gamma^{(f_2)}$ must be considered for $f = f_2\not=f_1$. Only
if $\Gamma^{(f_2)}$ is chosen appropriately, does the 
relation 
\begin{equation}
 {\chi_{\Gamma^{(f_1)}} \over \chi_{\Gamma^{(f_2)}} } \approx 
   \left( {f_1\over f_2} \right)^{-1+\eta(T)/2}
\end{equation}
hold for $f_1,f_2\to 0$. A naive choice would be 
$\Gamma^{(f_1)} = \Gamma^{(f_2)}$. As we now discuss,
this choice is not correct: different path families should 
be chosen for different values of $f$.

\begin{figure}[tb]
\centerline{\psfig{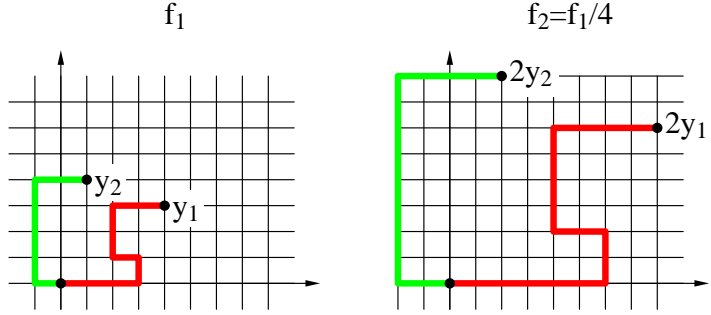}}
\vspace{2mm}
\caption{
One the left we report two paths connecting the origin to $y_1$ and $y_2$,
respectively. On the right, we report the corresponding paths
connecting the origin to $2y_1$ and $2y_2$.  The figure on the left
correspond to a frustration parameter $f = f_1$, that on the right to 
$f = f_2= f_1/4$.
}
\label{fig-Xf}
\end{figure}

To clarify this issue, let us imagine we are working in the continuum.
For each $f$, let us consider a family of paths 
$\Gamma^{(f)} = \{\Gamma^{(f)}_{x;y}\}$. Because of translation invariance,
we can limit ourselves to paths going from the origin to any point $y$.
These paths can be parametrised in terms of a function $X^{(f)}(t;y)$ such that 
$X^{(f)}(0;y) = 0$ for all $y$, $X^{(f)}(1;y) = y$. The path 
from the origin to $y$ is given by
\begin{equation}
   x = X^{(f)}(t;y) \qquad t\in [0,1].
\label{path-X}
\end{equation}
To determine the relation between $\Gamma^{(f_1)}$ and  $\Gamma^{(f_2)}$,
one should 
remember that $x/\xi$ should be kept fixed in the critical limit. 
Thus, we expect the path family to be invariant only if all lengths 
are expressed in terms of $\xi$. In other words, 
set $\bar{x} = x/\xi_f$, $\bar{y} = y/\xi_f$,
and rewrite (\ref{path-X}) as 
\begin{equation}
   \bar{x} = {1\over \xi_f} X^{(f)}(t;\bar{y}\xi_f) \qquad t\in [0,1],
\label{path-X2}
\end{equation}
where $\xi_f$ is the correlation length for the system with frustration 
parameter $f$. 
The natural requirement is therefore that the right hand side be independent
of $f$, that is 
\begin{equation}
   {1\over \xi_{f_2}} X^{(f_2)}(t;\bar{y}\xi_{f_2})  = 
   {1\over \xi_{f_1}} X^{(f_1)}(t;\bar{y}\xi_{f_1}) \; .
\end{equation}
Since we expect $\xi_f\sim f^{-1/2}$, we obtain the relation
\begin{equation}
X^{(f_2)}(t;r y) = r X^{(f_1)}(t;y), \qquad 
r = \left({f_1\over f_2}\right)^{1/2}.
\label{scal-rel-Gamma}
\end{equation}
In Fig.~\ref{fig-Xf} we report an example corresponding to $f_1 = 4 f_2$.
The paths from the origin to $y_1$ and $y_2$ which belong to $\Gamma^{(f_1)}$
completely fix the paths to $2y_1$ and $2y_2$ belonging to $\Gamma^{(f_2)}$.
Of course, on the lattice it is impossible to ensure 
(\ref{scal-rel-Gamma}) exactly. However, note that the relevant scale
is fixed by the correlation length and thus, violations at the level
of the lattice spacing are irrelevant in the critical limit.

\begin{figure}[tb]
\centerline{\psfig{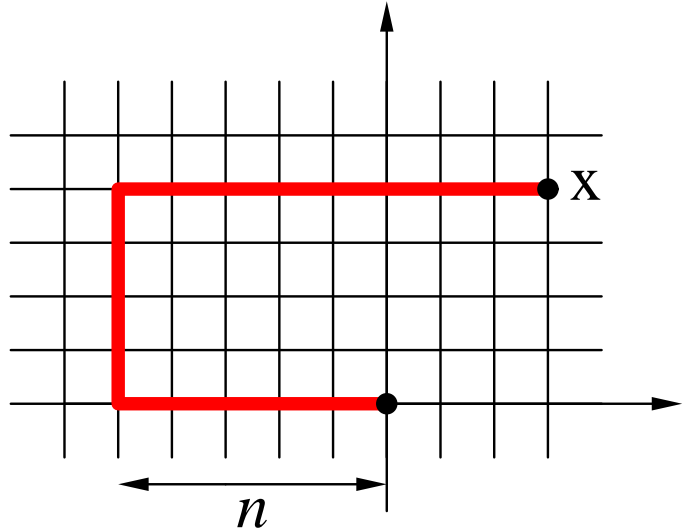}}
\vspace{2mm}
\caption{
The path connecting the origin to the point $x$ which belongs to the 
path family $\Gamma_n$.
}
\label{fig-Gamman}
\end{figure}

In the following we shall consider the path families 
$\Gamma_n \equiv  \{\Gamma_{n;0;x}\}$, which are specified by a non-negative
integer $n$. They are defined as follows 
(see Fig.~\ref{fig-Gamman}). The path $\Gamma_{n;0;x}$ connecting 
the origin to the point $x \equiv  (x_1,x_2)$ consists of three segments: 
the first one connects the origin to $(-n,0)$; the second one goes from
$(-n,0)$ to $(-n,x_2)$; the last one is horizontal, from 
$(-n,x_2)$ to point $x$. We indicate with $\chi_n(f)$ the corresponding 
susceptibilities and with $\xi_{n}(f)$ the corresponding
correlation lengths. These families of paths behave simply under the 
transformation (\ref{scal-rel-Gamma}). If we consider the 
path $\Gamma_{n;0;x}$ for $f = f_1$, the mapping 
(\ref{scal-rel-Gamma}) implies that, for $f = f_2$, one should consider the 
path $\Gamma_{rn;0;rx}$ between the origin and the point $rx$. 
This implies that, if we take the path family $\Gamma_n$ for $f = f_1$,
we must consider $\Gamma_{rn}$ for $f = f_2$. As a consequence,
$\chi_n$ and $\xi_n$ scale correctly only if we consider the limit
$n\to\infty$, $f\to 0$ at fixed $n f^{1/2}$. 
Thus, we predict the scaling behaviours 
\begin{eqnarray}
\chi_n &=& f^{-1+\eta(T)/2} F_\chi (n f^{1/2}), \nonumber \\
\xi_{n} &=& f^{-1/2} F_{\xi} (n f^{1/2}),
\label{scal-with-n}
\end{eqnarray}
where $F_\chi(x)$ and $F_\xi(x)$ are appropriate scaling functions.
In the next Section, we verify these predictions.

\section{Numerical results} \label{sec3}

We perform simulations for various values of $f=1/m$, $m$ integer, and $T$ in
the interval $T\le T_{KT}$, where $T_{KT}$ is the critical temperature of the 
$XY$ model, $T_{KT}=0.89294(8)$ \cite{Has-05}.
We consider finite lattices of size $L^2$, where 
$L$ is a multiple of $1/f$, and periodic boundary conditions for 
the spins.  Since we perform MC simulations in a gapped
phase, boundary conditions are expected to be irrelevant 
in the thermodynamic limit.  
Cluster algorithms cannot be used in the presence of frustration and thus 
we use an overrelaxed algorithm, which consists in performing
microcanonical and Metropolis updates. Predictions (\ref{xif}) and 
(\ref{chibeh}) hold in the thermodynamic limit, i.e. for
sufficiently large values of the ratio $L/\xi$,
where finite-size effects are negligible.
We find numerically that size effects are much smaller than our statistical 
errors for $Lf\gtrsim 3$.

In the simulations we choose the gauge
\begin{eqnarray}
&A_{\mvec{x}\mvec{y}}=  0 \quad  & {\rm if} \quad \mvec{y}= \mvec{x}+\hat{1}, \\
&A_{\mvec{x}\mvec{y}}=  {2 \pi f x_1}  \quad & 
   {\rm if} \quad \mvec{y}= \mvec{x}+\hat{2},
\nonumber
\label{lgau}
\end{eqnarray}
which is consistent with (\ref{deff}) and 
with $P_1(m) = P_2(m) = 1$, as long as $L$ is an integer 
multiple of $1/f$.  
With this gauge choice the computation of the susceptibilities 
$\chi_n$ and of the corresponding correlation lengths $\xi_n$ is quite simple. 
Indeed, $U[\Gamma_{n;x;y}] = 1$ for any $y$ 
if the first component of $x$ is $n$, 
i.e., if $x = (n,m)$, $m$ arbitrary.
Thus, if we choose $x = (n,m)$ in definition (\ref{chi-def}), 
we can compute $\chi_n$ without taking into account the phases $U_{xy}$. 
In practice, we have determined $\chi_n$ by using
\begin{equation}
\chi_n = {1\over L } \sum_{m}\sum_y 
\langle {\rm Re} \, \psi_{\mvec{(n,m)}}  \psi_{\mvec{y}}^* \rangle.
\label{chinMC}
\end{equation}
An analogous expression holds for the correlation lengths.

\begin{figure}[tb]
\centerline{\psfig{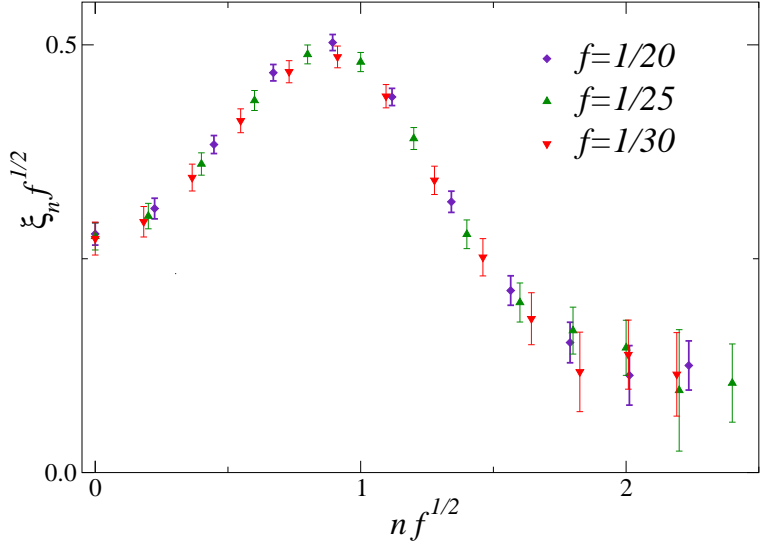}}
\vspace{2mm}
\caption{
Scaling plot for the correlation length $\xi_n$ at $T_{KT}$.
For each $f$ we report the data satisying $n \le 1/(2 f)$.
}
\label{fig:xiscal}
\end{figure}

\begin{figure}[tb]
\centerline{\psfig{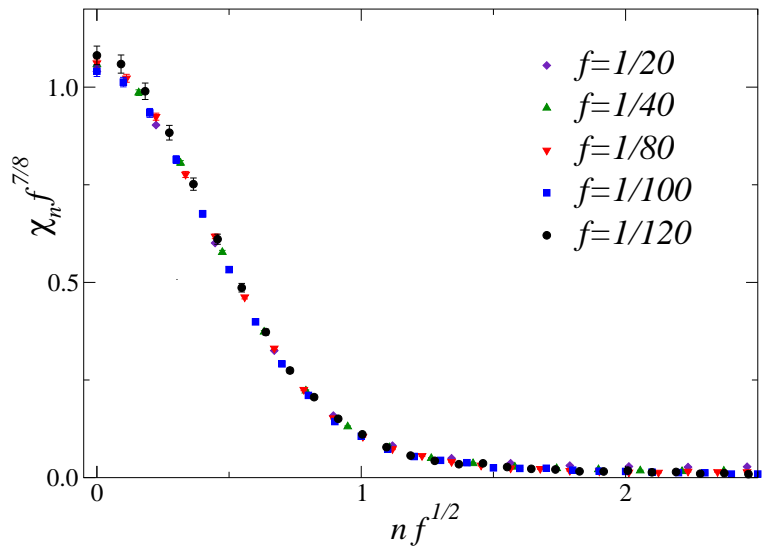}}
\vspace{2mm}
\caption{
Scaling plot for the susceptibilities $\chi_n$ at $T_{KT}$.
For each $f$ we report the data satisying $n \le 1/(2 f)$.
}
\label{fig:chiscal}
\end{figure}

In Figs.~\ref{fig:xiscal} and \ref{fig:chiscal} we plot the
correlation lengths $\xi_n$ and the susceptibilities $\chi_n$ at $T=T_{KT}$ 
for several values of $f$ and $n$. In this case $\eta(T)=1/4$ so that 
$\chi_n$ should scale as $f^{-7/8}$. It is easy to show 
that 
\begin{equation}
\chi_n = \chi_{n+1/f}, \qquad\qquad \chi_n = \chi_{1/f - n},
\label{chi-symm}
\end{equation}
 so that 
in (\ref{scal-with-n}) one must restrict oneself to data 
satisfying $0\le n\le 1/(2 f)$. The results reported in the figures
show the scaling behaviour (\ref{scal-with-n}) quite precisely, confirming 
the theoretical arguments. Note that the scaling function $F_\chi(x)$
apparently goes to zero as $x$ increases. This behaviour will be confirmed
below by the analysis of a non-gauge-invariant correlation function.

\begin{figure}[tb]
\centerline{\psfig{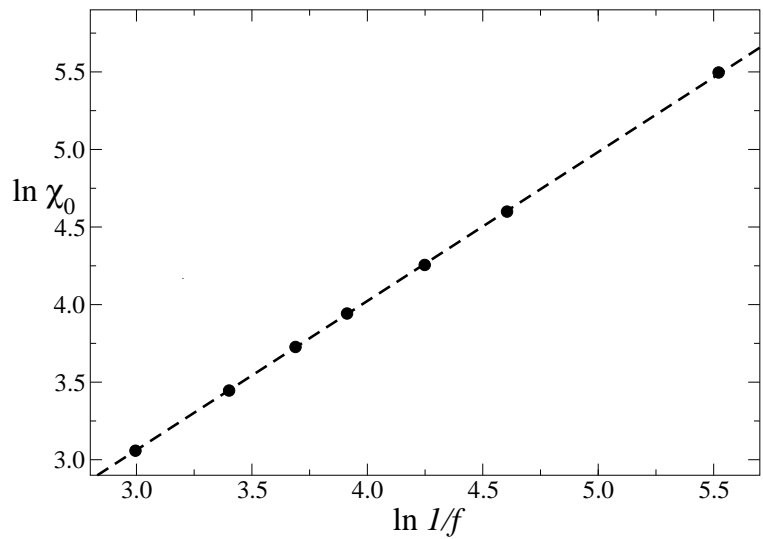}}
\vspace{2mm}
\caption{
Critical behaviour of 
$\chi_0$ vs $1/f$ at $T=0.4$.
The line is the results of a fit to $\chi=a f^{-\varepsilon}$,
which gives $\varepsilon=0.961(3)$, corresponding to
$\eta(T=0.4)=0.079(6)$. 
}
\label{chilt}
\end{figure}

Good agreement is also found at $T<T_{KT}$.  We
check the behaviour of $\chi_{n=0}$ (in this case, the same 
path family can be used for all values of $f$) up to $T=0.2$. 
At $T=0.2,0.3,0.4,0.5,0.8$, 
a fit of $\chi_0$ to $a f^{-1+\eta(T)/2}$ gives
$\eta=0.042(8),\,0.050(6),\,0.079(6),\,0.098(7),\,0.171(3)$. 
These results are in
substantial agreement with the leading spin-wave contibution $\eta=T/(2\pi)$,
and the MC estimates~\cite{etaest}
$\eta=0.036(3),\,0.052(5),\,0.074(6),\,0.100(8),\,0.19(2)$.
For example, in Fig.~\ref{chilt} we show the MC results for $\chi_0$
at $T=0.4$, together with the result of the fit. 
The data show a clear power-law behaviour in perfect agreement with 
(\ref{chibeh}).

We also investigated the critical behaviour of $\chi_{sq}$, which is expected
to scale as $f^{-3/4}$. For $1/f=40$, 60, 80, we obtain
$\chi_{sq} = 9.933(7)$, 13.630(23), 17.06(4), respectively. 
These results are fully consistent with the theoretical prediction.
Indeed, the product $f^{3/4} \chi_{sq}$ clearly converges to 
a constant as $f\to 0$ (corrections are expected to be proportional
to $1/\ln(1/f)$, as in the $XY$ model at $T_{KT}$): 
we have $f^{3/4} \chi_{sq} = 0.6245(5)$, 
0.6322(11), 0.6378(15) for the same values of $f$.

\begin{figure}[tb]
\centerline{\psfig{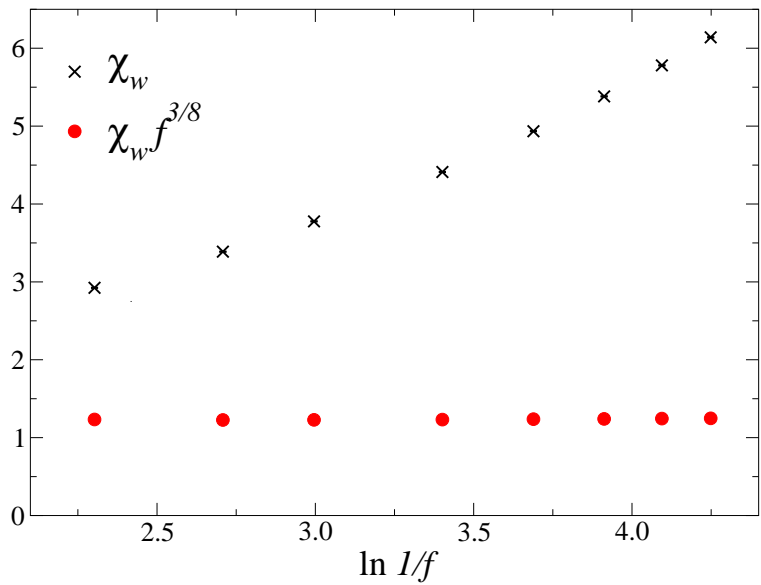}}
\vspace{2mm}
\caption{
MC results for the non-gauge-invariant susceptibility $\chi_w$ 
and for the product 
$f^{3/8}\chi_w$ vs $\ln 1/f$ at $T=T_{KT}$.
}
\label{chi-nongauge}
\end{figure}

Finally, we mention that correlation functions which are not gauge invariant
show a different behaviour. For example, one may consider the
susceptibility $\chi_w$ associated with the two-point function
$\langle {\rm Re} \, \psi_{\mvec{x}}
\psi_{\mvec{y}}^* \rangle$ in the gauge (\ref{lgau}):
\begin{equation}
   \chi_w = {1\over L^2} \sum_{x,y} \langle {\rm Re} \, \psi_{\mvec{x}}
\psi_{\mvec{y}}^* \rangle.
\end{equation}
At $T_{KT}$ it shows a
power-law behaviour $\chi_w\sim f^{-\varepsilon}$ as well,
but with a power
$\varepsilon\approx 0.39$, definitely different from the value
$0.875$ of the gauge-invariant definition.  
This result can be derived analytically. Indeed, we can rewrite
\begin{equation}
\chi_w = {1\over  L} \sum_{n=0}^{L-1} \chi_n,
\label{eq27}
\end{equation}
where $\chi_n$ is defined in (\ref{chinMC}).
Using the properties (\ref{chi-symm}) of the susceptibilities $\chi_n$,
(\ref{eq27}) can be rewritten as 
\begin{equation}
\chi_w \approx {2f} \sum_{n=0}^{1/(2f)} \chi_n.
\end{equation}
In this range of values of $n$, as is clear from Fig.~\ref{fig:chiscal},
we can use the scaling behaviour (\ref{scal-with-n}) and write
\begin{eqnarray}
\chi_w &\sim& f\times f^{-7/8} \int_0^{1/(2f)} dn\, F(nf^{1/2}) 
\nonumber \\
&\sim&
            f^{-3/8} \int_0^{1/(2f^{1/2})} dx\, F(x)\sim
            f^{-3/8} \int_0^\infty dx\, F(x).
\end{eqnarray}
Thus, provided that $F(x)$ is integrable (we already noted that 
the MC data for $\chi_n$ 
are consistent with $F(x)\to0$ as $x\to \infty$), we predict 
$\chi_w\sim f^{-3/8}=f^{-0.375}$, which is consistent with the 
MC data (see Fig.~\ref{chi-nongauge}). 

Note that the critical behaviour of $\chi_w$ depends on the chosen gauge.
If we use the gauge 
\begin{eqnarray}
&A_{\mvec{x}\mvec{y}}=  {-\pi f x_2} \quad  & 
      {\rm if} \quad \mvec{y}= \mvec{x}+\hat{1}, \\
&A_{\mvec{x}\mvec{y}}=  {\pi f x_1}  \quad & 
   {\rm if} \quad \mvec{y}= \mvec{x}+\hat{2},
\nonumber
\label{lgau2}
\end{eqnarray}
the susceptibility $\chi_w$ does not diverge and approaches a constant
as $f\to 0$.

In conclusion, we have shown that a small amount of uniform frustration 
(equivalently, a small uniform magnetic field)
destabilizes the line of fixed points that occur in the $XY$ model 
for $T \le  T_{KT}$. As soon as $f$ is different from zero, the system
becomes paramagnetic. The critical behaviour $\xi\sim f^{-1/2}$
can be predicted by simple Coulomb-gas and 
scaling arguments. Our numerical simulations fully confirm this prediction.
Also the scaling behaviour (\ref{chibeh}) for the magnetic susceptibilities 
is fully consistent with the numerical results.

\end{document}